\documentclass[cits]{PoS-pdf}
\title{Nonrelativistic effective field theory of unstable top}
\ShortTitle{Nonrelativistic effective field theory of unstable top}
\author{\speaker{Alexander A. Penin}$^{ab}$, Jan H. Piclum$^{cd}$\\
\llap{$^a$}Department of Physics, University of Alberta,\\
  Edmonton AB T6G 2J1, Canada\\
\llap{$^b$}Institut f{\"u}r Theoretische Teilchenphysik, Karlsruhe
  Institute of Technology (KIT),\\
  76128 Karlsruhe, Germany\\
\llap{$^c$}Institut f{\"u}r Theoretische Teilchenphysik und Kosmologie,
  RWTH Aachen,\\ 52056 Aachen, Germany\\
\llap{$^d$}Physik Department T31, Technische Universit{\"a}t
  M{\"u}nchen,\\ 85748 Garching, Germany\\
E-mail: \email{penin@ualberta.ca}, \email{jan.piclum@tum.de}}

\abstract{We develop a new nonrelativistic effective field theory of $\rho\!$NRQCD
\cite{Penin:2011gg} to describe the finite lifetime effects in the threshold production
of top quark-antiquark pairs. The theory is based on the expansion in
a parameter $\rho=1-m_W/m_t$ characterizing the dynamics of the top-quark decay.
Within this framework we compute the nonresonant contribution to the
total cross section of the top quark-antiquark threshold production in
electron-positron annihilation up to the next-to-next-to-leading order.
Our method naturally resolves the problem of spurious divergences in the analysis
of the unstable top production.}
\FullConference{Loops and Legs in Quantum Field Theory, Wernigerode, Germany, 15-20 April 2012}

\newcommand{\iep}{i\varepsilon}
\newcommand{\qe}{Q_e}
\newcommand{\qt}{Q_t}

\renewcommand{\ae}{a_e}
\newcommand{\ve}{v_e}

\newcommand{\vt}{v_t}

\newcommand{\cw}{c_w}
\newcommand{\sw}{s_w}
\newcommand{\vecp}{\vec{p}\,}

\begin{document}

\section{Introduction}

The threshold production of top quark-antiquark pairs at a future linear
collider may provide us with the most accurate information on the top-quark mass
and couplings crucial for our understanding of electroweak symmetry breaking and
mass generation mechanism~\cite{Martinez:2002st}. The nonrelativistic
top-antitop pair is the cleanest quarkonium-like  system and its theoretical
description is entirely based on the first principles of QCD
\cite{Fadin:1987wz,Fadin:1988fn,Strassler:1990nw}. Over the last decade a
significant progress has been achieved in the analysis of the higher order
perturbative and relativistic corrections  in the strong coupling constant
$\alpha_s$ and the heavy-quark velocity $v$ to the threshold cross section.
Sizable next-to-next-to-leading order (NNLO) corrections have been found
by several groups
\cite{Hoang:1998xf,Melnikov:1998pr,Penin:1998zh,Beneke:1999qg,Penin:1998mx,Nagano:1999nw}
that stimulated the study of the higher orders of perturbation theory. Currently
a large part of the third order corrections is available
\cite{Kniehl:1999ud,Kniehl:1999mx,Kniehl:2002br,Penin:2002zv,Kniehl:2002yv,
Hoang:2003ns,Penin:2005eu,Beneke:2005hg,Marquard:2006qi,Beneke:2007gj,
Beneke:2007pj,Beneke:2008cr,Marquard:2009bj, Anzai:2009tm,Smirnov:2009fh} and
the N$^3$LO analysis is likely to be completed in the foreseeable future.
Moreover the  higher order logarithmically enhanced corrections have been
resummed  through the effective theory renormalization group methods
\cite{Hoang:2000ib,Penin:2004ay,Pineda:2006ri}.

At the same time much less attention has been paid to the analysis of the
effects related to the instability of the top quark
\cite{Bigi:1991mi,Melnikov:1993np,Penin:1998ik,Hoang:2004tg}. The width of the
electroweak top-quark  decay $t\to W^+b$, $\Gamma_t\approx 1.5$~GeV, is
comparable to the binding energy of a would-be toponium ground state and has a
dramatic effect on the threshold production. It serves as an infrared cutoff,
which makes the process perturbative in the whole threshold region, and smears
out all the Coulomb-like resonances below the threshold leaving  a single well
pronounced peak in the cross section. The standard prescription in the analysis
of the unstable top-quark production consists of the complex shift  $E\to
E+i\Gamma_t$, where $E$ is the top-antitop pair energy counted from the
threshold  \cite{Fadin:1987wz}.  Though this procedure incorporates  the
dominant effect of the finite top-quark width, it does not fully account for
nonresonant processes like $e^+e^-\to t W^-\bar{b}$, $e^+e^-\to b W^+\bar{t}$ or
$e^+e^-\to W^+W^- b\bar{b}$ where the intermediate top quark is not on its
(complex) mass shell.  Such processes {\em cannot be distinguished} from the
resonant $t\bar{t}$ production, which has the same final states due to the
top-quark instability.

Recently different approaches have been suggested to refine the analysis of  the
finite width effect \cite{Penin:2011gg,Hoang:2010gu,Beneke:2010mp}. The method
of ref.~\cite{Penin:2011gg} is based on a new nonrelativistic effective theory
which we name $\rho\!$NRQCD. In this proceeding we
outline its main idea and present the results obtained within this framework.

\section{Finite width effect beyond the complex energy shift}
\begin{figure}[t]
  \begin{center}
    \includegraphics[width=0.7\textwidth]{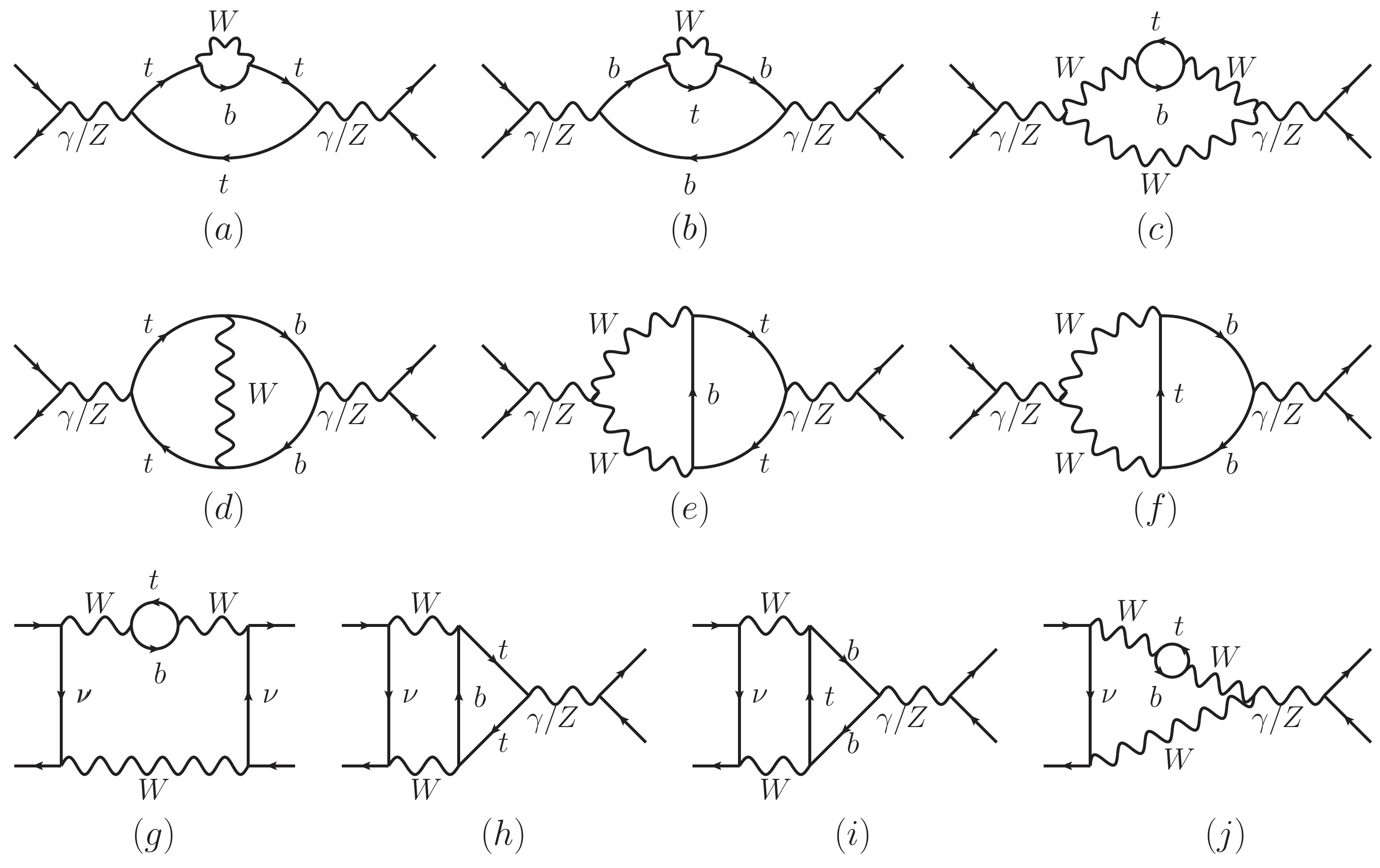}
  \end{center}
  \caption{\label{fig::dias}$e^+e^-$ forward scattering diagrams
    containing $b W^+\bar{t}$ and $t W^-\bar{b}$ cuts.}
\end{figure}

In the Born approximation the total cross section of top-antitop production
in electron-positron annihilation is related through the optical theorem to the
imaginary part of the one-loop forward scattering amplitude.
The corresponding expression for the normalized cross
section $R ={\sigma(e^+e^-\to t\bar t)/\sigma_0}$, $\sigma_0=4\pi\alpha^2/(3s)$,
in the threshold region $s\approx 4m_t^2$ can be obtained by the standard
nonrelativistic expansion of the top-quark vertices and propagators in $v$
and reads
\begin{eqnarray}
R^{Born}_{res}= \left[ \qe^2\qt^2 + \frac{2\qe\qt\ve\vt}{1-x_Z} +
\frac{(\ae^2+\ve^2)\vt^2}{(1-x_Z)^2} \right]{6\pi N_c\over m_t^2}\,
{\rm Im}[G_0(0,0,E+\iep)]
+\ldots\,,
\label{res}
\end{eqnarray}
where the ellipsis stands for the relativistic corrections, $Q_f$ is the
electric charge of fermion $f$ in units of the positron charge, $N_c=3$ is the
number of colors, and $x_Z=m_Z^2/(4m_t^2)$ with the $Z$-boson mass $m_Z$. The
couplings of fermion $f$ to the $Z$-boson are
\begin{equation}
v_f = \frac{I^3_{w,f} - 2\sw^2Q_f}{2\sw\cw}\,, \qquad a_f =
\frac{I^3_{w,f}}{2\sw\cw} \,,
\end{equation}
where $I^3_{w,f}$ is the third component of the fermion's weak isospin and $\sw$
($\cw$) is the sine (cosine) of the weak mixing angle. Note that only the vector
coupling of the top quark gives the leading order contribution and the axial
coupling is suppressed by an additional power of $v$. The last factor in
eq.~(\ref{res}) is
\begin{eqnarray}
G_0(0,0,E)=\int {{\rm
d}^{d-1}\vecp\over (2\pi)^{d-1}}{m_t\over{\vecp^2-m_tE}}
=-{m_t^2\over 4\pi}\sqrt{{-E\over m_t}}\,,
\label{g0}
\end{eqnarray}
which is nothing but the Green's function of the free Schr\"odinger equation at
the origin. Formally the integral in eq.~(\ref{g0}) is linearly divergent  but
the divergent part is real and does not contribute to the cross section. To
handle the divergence we use dimensional regularization with $d=4-2\epsilon$,
where the integral~(\ref{g0}) is finite even for $\epsilon =0$.
The strong interaction has a significant impact on the threshold cross section.
Close to threshold when $v\sim \alpha_s$ the Coulomb effects become
nonperturbative and have to be resummed to all orders in $\alpha_s$ by
substituting eq.~(\ref{g0}) with the full Coulomb Green's function
\begin{eqnarray}
G_C(0,0;E)&=&G_0(0,0;E)+G_1(0,0;E)-{C_F\alpha_sm_t^2\over
4\pi}\left[\Psi\left(1- {C_F\alpha_s\over
2}\sqrt{m_t\over -E} \right)+\gamma_E\right]\,,
\nonumber
\\
&&\label{gc}
\end{eqnarray}
where $\Psi$ is the logarithmic derivative of the Gamma function and
$C_F=4/3$. The one-gluon exchange contribution $G_1(0,0;E)$ is ultraviolet
divergent. Again for stable top quarks the divergent part is real and does not
contribute to eq.~(\ref{res}). In the $\overline{\rm MS}$ subtraction scheme
this term reads
\begin{eqnarray}
G_1(0,0;E)&=&-{C_F\alpha_sm_t^2\over
8\pi}\left[\ln\left({-m_tE\over \mu^2}\right)-1+2\ln{2}\right]\,.
\label{g1}
\end{eqnarray}
Let us now consider the top-quark decay. Every decay process is suppressed by
the electroweak coupling constant $\alpha_{ew}$. We adopt the standard power
counting rules $\alpha_s\sim v$, $\alpha_{ew}\sim v^2$. Thus to NNLO if the
top quark decays the antiquark may be treated as a stable particle and vice
versa.  The dominant effect of the top-quark instability is related to the
imaginary part of its mass operator in diagram~\ref{fig::dias}$(a)$.
In the massless bottom quark approximation and with the off-shell momentum $p$
the mass operator reads
\begin{equation}
 {\rm Im}[\Sigma^{(0)}(p^2)]={G_F\over
16\pi\sqrt{2}}\,{p^3}\left(1+2{m^2_W\over p^2}\right)\left(1-{m^2_W\over
p^2}\right)^2
\theta(p^2-m_W^2)\,,
\label{sigma0}
\end{equation}
where $G_F$ is the Fermi constant and we use the approximation $V_{tb}= 1$.
Close to the mass shell one has $p^2=m_t^2-2(\vecp^2-m_tE)+\ldots$, where
$\vecp$ is the spatial momentum of the top quark, and the mass operator can be
expanded in $z=(\vecp^2-m_tE)/m_t^2\ll 1$
\begin{eqnarray}
 {\rm
Im}[\Sigma^{(0)}(z)]&=&{\Gamma_t\over 2}\left(1-{4z\over
(1-x^2)}+{4z^2\over
(1-x^2)^2}\right)\theta(1-x^2-2z)+\ldots
\nonumber
\\
&=& {\Gamma_t\over 2}-{\Gamma_t\over 2}\!\left[\theta(x^2+2z-1)+\left({4z\over
(1-x^2)}-{4z^2\over
(1-x^2)^2}\right)\theta(1-x^2-2z)\right]\!+\!\ldots\,.
\nonumber
\\
\label{sigma0os}
\end{eqnarray}
where $x=m_W/m_t$, $\Gamma_t=\Gamma^{(0)}_t+{\cal O}(\alpha_s)$ and
\begin{equation}
\Gamma^{(0)}_t={G_Fm_t^3\over 8\pi\sqrt{2}}(1+2x^2)(1-x^2)^2\,,
\label{gamma0}
\end{equation}
is the leading order top-quark electroweak width. Note that in the
expansion~(\ref{sigma0os}) we consider $1-x=\rho$ to be of the same order of
magnitude as $z$. The first term in the last line of eq.~(\ref{sigma0os})
describes the standard shift of the pole position of the top quark  propagator
into the unphysical sheet of the complex energy plane characteristic for
unstable particles. After Dyson resummation it replaces the argument of
eq.~(\ref{g0}) by $E+i\Gamma_t$, which is the original prescription of
ref.~\cite{Fadin:1987wz}. Eq.~(\ref{sigma0os}), however, contains the remainder which
also contributes to the imaginary part of the forward scattering amplitude.
Since the remainder vanishes for on-shell top quark, it represents the
nonresonant process $e^+e^-\to b W^+\bar{t}$ or $e^+e^-\to t W^-\bar{b}$.
Direct evaluation of the nonresonant contribution is technically complex
as the problem involves a large number of scales characterizing both
the top-quark threshold dynamics and the dynamics of the top-quark decay.
Such a problem, however, is an ideal application of the effective field
theory approach.

\section{Nonrelativistic effective theory of unstable top}
In the nonresonant contribution the integral over the virtual momentum $\vecp$
is saturated by the region $|\vecp|\sim \rho^{1/2} m_t$ where the argument of
the $\theta$-function in eq.~(\ref{sigma0os}) vanishes. The main idea of our
approach is that if $\rho$ is considered as a small parameter this momentum
region corresponds to a nonrelativistic top quark with the energy $p_0-m_t\sim
\vecp^2/m_t \sim \rho m_t$.   It is equivalent to the top quark momentum
scaling in the standard potential nonrelativistic QCD (pNRQCD)
\cite{Pineda:1997bj,Brambilla:1999xf} with the heavy quark velocity $v$
replaced by $\rho^{1/2}$. Thus beside the hard scale $m_t$, the soft  scale
$vm_t$, and the ultrasoft scale   $v^2m_t$  characterizing  the top-quark
threshold dynamics we have $\rho$-soft  scale $\rho^{1/2}m_t$ and
$\rho$-ultrasoft  scale $\rho m_t$ characterizing  the dynamics of the
top-quark decay. In full analogy with the pNRQCD the new scales give rise to the
$\rho$-soft and $\rho$-ultrasoft modes as well as to the $\rho$-potential  modes
with the nonuniform scaling of energy and three-momentum $p_0-m_t\sim
\rho m_t,~|\vecp|\sim \rho^{1/2}m_t$. The $\rho\!$NRQCD is a  new nonrelativistic
effective  theory which incorporates  these modes. To disentangle the ordinary
pNRQCD and the new $\rho\!$NRQCD modes we impose the scale  hierarchy
$vm_t\ll\rho^{1/2} m_t\ll m_t$ or $v\ll \rho^{1/2} \ll1$. The cross section is
then constructed as a series in the scale ratios.

\begin{figure}[t]
  \begin{center}
    \includegraphics[width=0.5\textwidth]{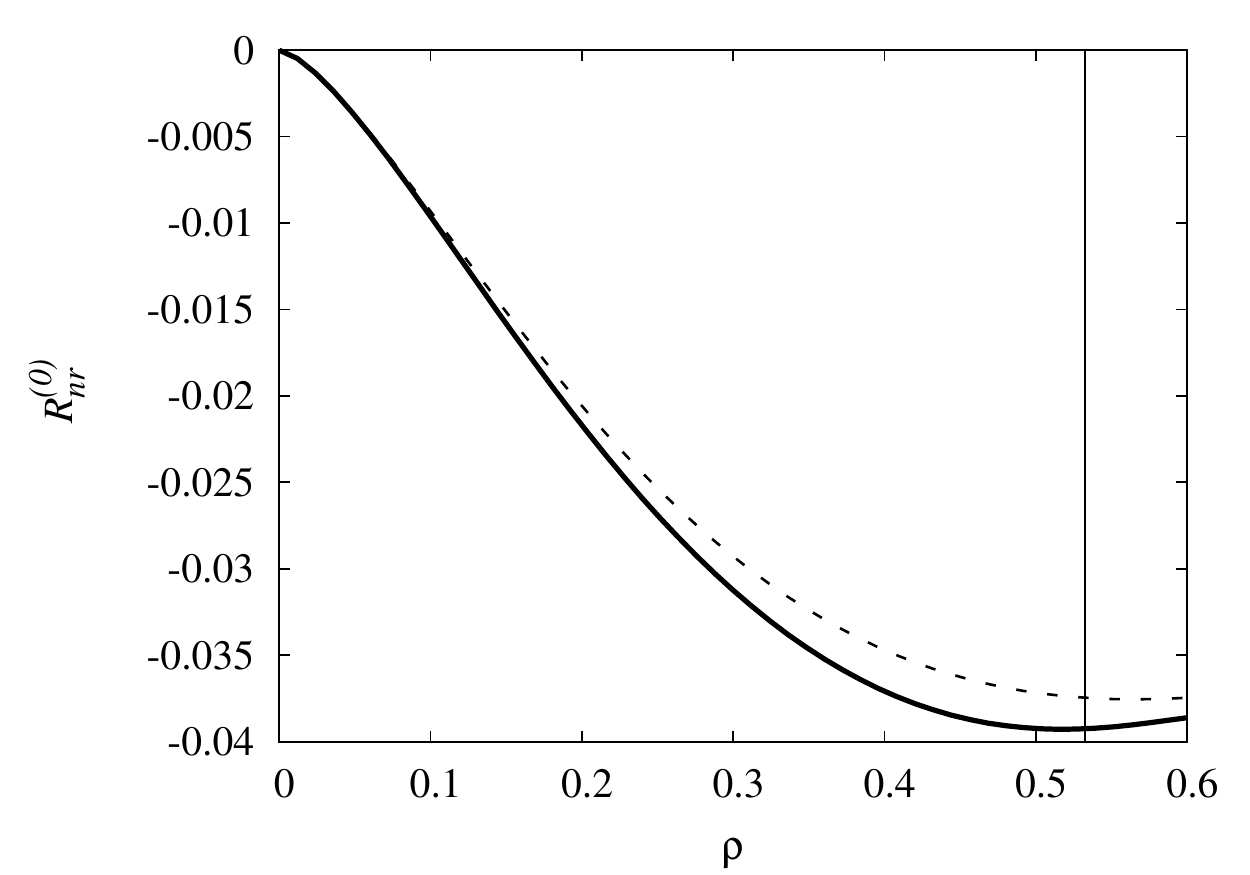}
  \end{center}
  \caption{\label{fig::dR1} The NLO nonresonant
  contribution to the cross section
(solid curve)
    and the leading order approximation
(dashed line). The
    vertical line marks the physical value of $\rho$.}
\end{figure}

Eq.~(\ref{sigma0os}) does not give the full nonresonant contribution and one has
to take into account all diagrams of this order with the $b W^+\bar{t}$ or $t
W^-\bar{b}$ cuts given in figure~\ref{fig::dias}. The calculation is
significantly simplified within the nonrelativistic effective theory where all
the propagators which are off-shell by the amount $m_t$ collapse, giving rise to
new effective theory vertices. From the practical point of view, however, it is
more convenient to directly expand the full theory Feynman integrals in $\rho$
through the expansion by regions~\cite{Beneke:1997zp,Smirnov:2002pj} than to use
the effective theory Feynman rules. The contribution comes from a single region
where the top quark and $W$-boson are $\rho$-potential and the bottom quark is
$\rho$-ultrasoft and one may neglect all the terms suppressed by the ratio
$v^2/\rho$, {\it i.e.} put $E=0$. We found that beside
diagram~\ref{fig::dias}$(a)$ only diagram~\ref{fig::dias}$(g)$ gives a leading
order contribution in  $\rho$.  For the total nonresonant contribution to the cross
section at the leading order in $\rho$ we obtain a simple analytical
result
\begin{eqnarray}
R_{nr}&=&- {8N_c\over \pi \rho^{1/2}}{\Gamma_t\over m_t} \Bigg[
\left(\qe^2\qt^2
+ \frac{2\qe\qt\ve\vt}{1-x_Z} +\frac{(\ae^2+\ve^2)\vt^2}{(1-x_Z)^2} \right)
\nonumber
\\
&&\left.
-{1\over \sw^4} \left({17\over 48}-{9\sqrt{2}\over
    32}\ln\left(1+\sqrt{2}\right)\right)+{\cal O}(\rho,\alpha_s)\right]\,.
\label{rnrlo}
\end{eqnarray}
It is suppressed with respect to the leading resonant contribution~(\ref{res}) and
according to the power counting rules represents a NLO contribution to the total cross
section.

\subsection{Relativistic and perturbative corrections}
\label{sec::rhoexp}

Eq.~(\ref{rnrlo}) gives the nonresonant contribution in the leading order of the
nonrelativistic expansion in  $\rho$ and perturbative expansion in $\alpha_s$.
In the leading order in $\alpha_s$ it is straightforward to compute a sufficient
number of terms of the expansion in $\rho$ of the diagrams in Fig.~\ref{fig::dias} to
ensure very good accuracy  of the approximation for the physical value
$\rho\approx 0.53$ \cite{Penin:2011gg}. The Pad\'{e}-improved series converges
to the result of \cite{Beneke:2010mp} obtained without the expansion. The
result for the NLO contribution after summation of the series is plotted  in
figure~\ref{fig::dR1} as function of $\rho$ along with  the leading order
term~(\ref{rnrlo}). The latter turns out to be a good approximation in the whole
interval $0<\rho<0.6$ and deviates from the total result by less than $5\%$ at
the physical value of $\rho$.

\begin{figure}[t]
  \begin{center}
    \includegraphics[width=0.85\textwidth]{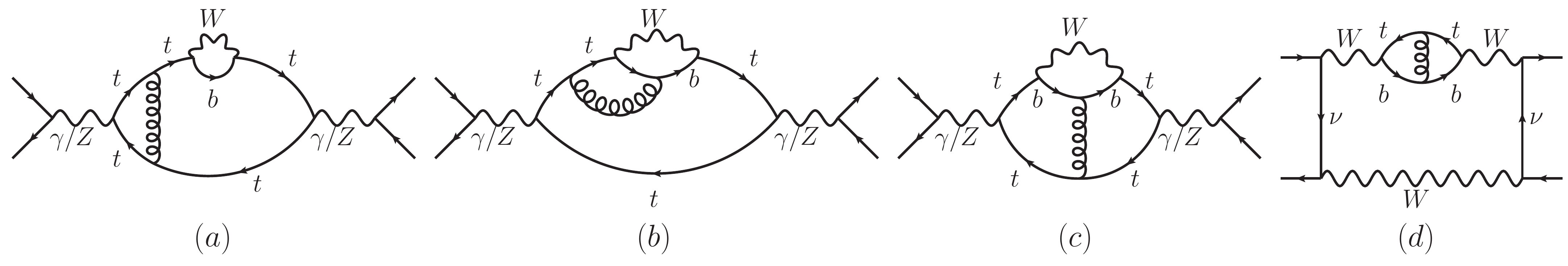}
  \end{center}
  \caption{\label{fig::ascorr}$e^+e^-$ forward scattering diagrams
    contributing to the $\mathcal{O}(\alpha_s)$ correction in the leading
    order in $\rho$. The curly lines denote gluons.}
\end{figure}

In the leading order in $\rho$ the $\alpha_s$ corrections are obtained by the
gluon dressing of diagrams~\ref{fig::dias}$(a)$ and~\ref{fig::dias}$(g)$ shown
in figure~\ref{fig::ascorr}.  The  $\rho\!$NRQCD scaling and power counting
rules identify the relevant regions of virtual momentum. In particular, the
diagram~\ref{fig::ascorr}$(a)$ gets contributions from the hard, potential, and
$\rho$-potential regions of gluon momentum, the diagrams~\ref{fig::ascorr}$(b)$
and  \ref{fig::ascorr}$(d)$ get contributions from the hard and $\rho$-soft
regions, while the diagram~\ref{fig::ascorr}$(c)$ vanishes in the leading order
of the nonrelativistic expansion \cite{Melnikov:1993np}. By adding up the
contributions of all the regions one gets the ${\cal O}(\alpha_s)$ correction to
eq.~(\ref{rnrlo}). It represents the  NNLO nonresonant contribution to the total
cross section in the leading order in $\rho$  and can be obtained in a closed
analytical form
\cite{Penin:2011gg}
\begin{eqnarray}
R^{(1)}_{nr} &=& \frac{N_c C_F\alpha_s}{\pi^2 \rho^{1/2}}\,
\frac{\Gamma_t}{m_t} \bigg\{ \left[ \qe^2\qt^2 +
\frac{2\qe\qt\ve\vt}{1-x_Z} +
\frac{(\ae^2+\ve^2)\vt^2}{(1-x_Z)^2} \right]
\nonumber\\
&&\times\left[\left(3\ln\left({\sqrt{E^2+\Gamma_t^2}\over \rho m_t}\right)+\frac{3}{2}+6\ln2\right){\pi^2\over\rho^{1/2}}
+\left(18 + 24\ln2 \right)\right]
\nonumber
\\
&& + \frac{1}{\sw^4} \bigg[ \frac{22}{3} + \frac{17\pi^2}{6} -
\frac{17}{2} \ln2 + \left(2 - 3\pi^2 + 9\ln2 \right) \frac{3\sqrt{2}}{4}
\ln\left(1+\sqrt{2}\right)
\nonumber
\\
&& - \frac{27\sqrt{2}}{8} \left( \ln^2\left(1+\sqrt{2}\right) +
\mathrm{Li}_2\left(2\sqrt{2}-2\right) \right) \bigg]
+{\cal O}(\rho)\bigg\}\,.
\label{r1res}
\end{eqnarray}
Eq.~(\ref{r1res}) includes  $1/\rho^{1/2}$ and $\ln(\rho/v^2)$  enhanced
contributions. The $1/\rho^{1/2}$ power enhanced term is due to the Coulomb
singularity in full analogy with the $1/v$ enhancement of the Coulomb gluon
exchange in pNRQCD. An important difference with respect to pNRQCD is that since
$\alpha_s/\rho^{1/2}\ll 1$ one does not need to resum the $\rho$-Coulomb
corrections to all orders.  The $\ln(\rho/v^2)$ term represents a new type of
the large logarithms in the theory of top-quark threshold production. It
originates from the logarithmic integral between the potential and $\rho$-potential
momenta. Within the expansion by regions it shows up through the
infrared divergence of the $\rho$-potential region and the ultraviolet divergence
of the ordinary potential region. In the total result the divergences
cancel each other leaving the logarithm of the scale ratio. Note that the contribution
of the potential region alone is ultraviolet divergent. This is exactly the origin
of the spurious divergences in pNRQCD description of the unstable top production
which does not include the $\rho$-potential region \cite{Bigi:1991mi,Penin:1998ik}.
Thus, the problem of the spurious divergences is naturally solved in  $\rho\!$NRQCD.

\subsection{Numerical results}
\label{sec::res}

\begin{figure}[t]
  \begin{center}
    \includegraphics[width=0.5\textwidth]{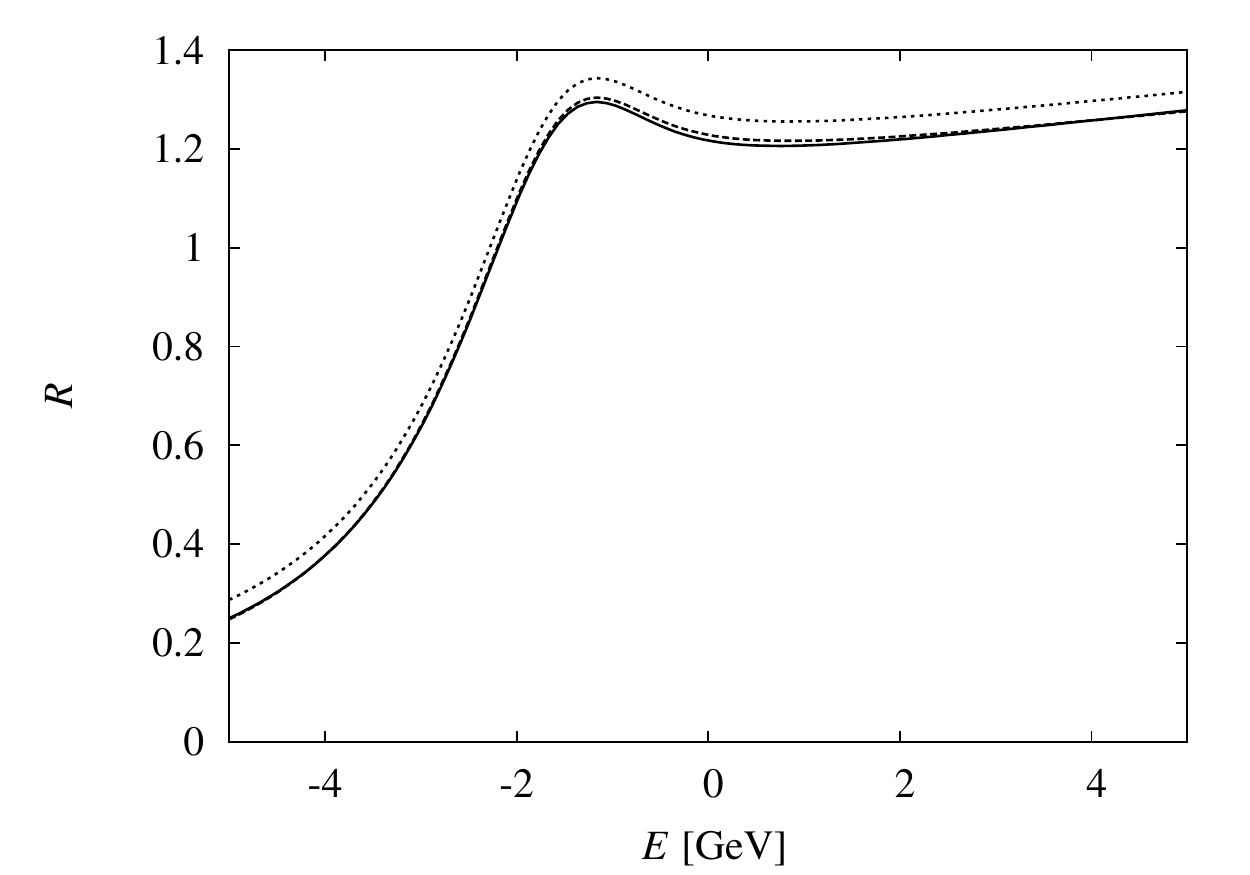}
  \end{center}
  \caption{\label{fig::full} The normalized cross section of top-antitop
  production in electron-positron annihilation as  function of the
  energy counted from the  threshold. The dotted curve represents the leading
  order pNRQCD Coulomb approximation. The dashed curve includes the
  NLO nonresonant contribution  and the solid curve includes the NNLO
  nonresonant correction as well. No strong coupling corrections to the
  resonance contribution are included.}
\end{figure}

The numerical effect of the nonresonant contribution on the total
threshold cross section is shown in figure~\ref{fig::full}. In this plot we use
the leading order pNRQCD approximation for the resonance contribution
corresponding to the Coulomb Green's function~(\ref{gc}) with the strong
coupling constant normalized at the soft scale $\mu_s=\alpha_s(\mu_s)C_Fm_t$.
The NLO contribution  is negative and amounts to about 3.1\% of the leading
cross section above the threshold and competes with the LO contribution
below the resonance region.  The NNLO nonresonant contribution amounts
to $-0.9\%$ at the threshold.

\section{Summary}
\label{sec::con}
In this paper we presented the $\rho\!$NRQCD,  a new nonrelativistic effective
field theory which systematically accounts for the top-quark instability in the
threshold top quark-antiquark pair production.  The theory is  based on the
nonrelativistic expansion in the  parameter $\rho=1-m_W/m_t$. It  is optimized
for high-order calculations and solves the problem of the spurious divergences
of the standard pNRQCD approach. Within this framework we obtain the  NLO
nonresonant contribution to the total threshold cross section of the top
quark-antiquark pair production in electron-positron annihilation  and extended
the analysis to the NNLO ${\cal O}(\alpha_s)$ nonresonant contribution which is
computed to the leading order in $\rho$.


\begin{thebibliography}{99}

\bibitem{Penin:2011gg}
  A.~A.~Penin and J.~H.~Piclum,
  {\it Threshold production of unstable top},
  JHEP {\bf 1201} (2012) 034
  [arXiv:1110.1970 [hep-ph]].

\bibitem{Martinez:2002st}
  M.~Martinez, R.~Miquel,
  {\it Multiparameter fits to the $t\bar t$ threshold observables at a future
$e^+e^-$ linear collider},
  Eur.\ Phys.\ J.\  C {\bf 27 } (2003)  49
  [hep-ph/0207315].

\bibitem{Fadin:1987wz}
  V.~S.~Fadin, V.~A.~Khoze,
  {\it Threshold Behavior of Heavy Top Production in $e^+e^-$ Collisions},
  JETP Lett.\  {\bf 46 } (1987)  525.

\bibitem{Fadin:1988fn}
  V.~S.~Fadin, V.~A.~Khoze,
  {\it Production of a pair of heavy quarks in $e^+e^-$ annihilation in the
threshold region},
  Sov.\ J.\ Nucl.\ Phys.\  {\bf 48 } (1988)  309.

\bibitem{Strassler:1990nw}
  M.~J.~Strassler, M.~E.~Peskin,
  {\it The Heavy top quark threshold: QCD and the Higgs},
  Phys.\ Rev.\  D {\bf 43 } (1991)  1500.

\bibitem{Hoang:1998xf}
  A.~H.~Hoang, T.~Teubner,
  {\it Top quark pair production at threshold: Complete next-to-next-to-leading
order relativistic corrections},
  Phys.\ Rev.\  D {\bf 58 } (1998)  114023
  [hep-ph/9801397].

\bibitem{Melnikov:1998pr}
  K.~Melnikov, A.~Yelkhovsky,
  {\it Top quark production at threshold with $O(\alpha_s^2)$ accuracy},
  Nucl.\ Phys.\  B {\bf 528 } (1998)  59
  [hep-ph/9802379].

\bibitem{Penin:1998zh}
  A.~A.~Penin and A.~A.~Pivovarov,
  {\it Next-to-next-to leading order vacuum polarization function of heavy quark near threshold and sum rules for b anti-b system},
  Phys.\ Lett.\ B {\bf 435} (1998) 413
  [hep-ph/9803363].

\bibitem{Beneke:1999qg}
  M.~Beneke, A.~Signer, V.~A.~Smirnov,
  {\it Top quark production near threshold and the top quark mass},
  Phys.\ Lett.\  B {\bf 454 } (1999)  137
  [hep-ph/9903260].

\bibitem{Penin:1998mx}
  A.~A.~Penin, A.~A.~Pivovarov,
  {\it Analytical results for $e^+e^-\to t\bar t$  and $\gamma\gamma\to t\bar t$
observables near the threshold up to the next-to-next-to leading order of
NRQCD},
  Phys.\ Atom.\ Nucl.\  {\bf 64 } (2001)  275
  [hep-ph/9904278].

\bibitem{Nagano:1999nw}
  T.~Nagano, A.~Ota, Y.~Sumino,
  {\it $O(\alpha_s^2)$  corrections to $e^+e^-\to t\bar t$  total and
differential cross-sections near threshold},
  Phys.\ Rev.\  D {\bf 60 } (1999)  114014
  [hep-ph/9903498].

\bibitem{Kniehl:1999ud}
  B.~A.~Kniehl, A.~A.~Penin,
  {\it Ultrasoft effects in heavy quarkonium physics},
  Nucl.\ Phys.\  B {\bf 563 } (1999)  200
  [hep-ph/9907489].

\bibitem{Kniehl:1999mx}
  B.~A.~Kniehl, A.~A.~Penin,
  {\it Order $\alpha_s^3 \ln^2 (1 /\alpha_s)$ corrections to heavy quarkonium
creation and annihilation},
  Nucl.\ Phys.\  B {\bf 577 } (2000)  197
  [hep-ph/9911414].

\bibitem{Kniehl:2002br}
  B.~A.~Kniehl, A.~A.~Penin, V.~A.~Smirnov, M.~Steinhauser,
  {\it Potential NRQCD and heavy quarkonium spectrum at
next-to-next-to-next-to-leading order},
  Nucl.\ Phys.\  B {\bf 635 } (2002)  357
  [hep-ph/0203166].

\bibitem{Penin:2002zv}
  A.~A.~Penin, M.~Steinhauser,
  {\it Heavy quarkonium spectrum at $O(\alpha^5_s m_q)$ and bottom / top quark
mass determination},
  Phys.\ Lett.\  B {\bf 538 } (2002)  335
  [hep-ph/0204290].

\bibitem{Kniehl:2002yv}
  B.~A.~Kniehl, A.~A.~Penin, M.~Steinhauser, V.~A.~Smirnov,
  {\it Heavy quarkonium creation and annihilation with $\alpha_s^3 \ln (1
/\alpha_s)$  accuracy},
  Phys.\ Rev.\ Lett.\  {\bf 90 } (2003)  212001
  [hep-ph/0210161].

\bibitem{Hoang:2003ns}
  A.~H.~Hoang,
  {\it Three loop anomalous dimension of the heavy quark pair production current
in nonrelativistic QCD},
  Phys.\ Rev.\  D {\bf 69 } (2004)  034009
  [hep-ph/0307376].

\bibitem{Penin:2005eu}
  A.~A.~Penin, V.~A.~Smirnov, M.~Steinhauser,
  {\it Heavy quarkonium spectrum and production/annihilation rates to order
$\beta_0^3\alpha_s^3$},
  Nucl.\ Phys.\  B {\bf 716 } (2005)  303
  [hep-ph/0501042].

\bibitem{Beneke:2005hg}
  M.~Beneke, Y.~Kiyo, K.~Schuller,
  {\it Third-order Coulomb corrections to the S-wave Green function, energy
levels and wave functions at the origin},
  Nucl.\ Phys.\  B {\bf 714 } (2005)  67
  [hep-ph/0501289].

\bibitem{Marquard:2006qi}
  P.~Marquard, J.~H.~Piclum, D.~Seidel, M.~Steinhauser,
  {\it Fermionic corrections to the three-loop matching coefficient of the
vector current},
  Nucl.\ Phys.\  B {\bf 758 } (2006)  144
  [hep-ph/0607168].

\bibitem{Beneke:2007gj}
  M.~Beneke, Y.~Kiyo, K.~Schuller,
  {\it Third-order non-Coulomb correction to the S-wave quarkonium wave
functions at the origin},
  Phys.\ Lett.\  B {\bf 658 } (2008)  222
  [arXiv:0705.4518 [hep-ph]].

\bibitem{Beneke:2007pj}
  M.~Beneke, Y.~Kiyo, A.~A.~Penin,
  {\it Ultrasoft contribution to quarkonium production and annihilation},
  Phys.\ Lett.\  B {\bf 653 } (2007)  53
  [arXiv:0706.2733 [hep-ph]].

\bibitem{Beneke:2008cr}
  M.~Beneke, Y.~Kiyo,
  {\it Ultrasoft contribution to heavy-quark pair production near threshold},
  Phys.\ Lett.\ B {\bf 668 } (2008)  143
  [arXiv:0804.4004 [hep-ph]].

\bibitem{Marquard:2009bj}
  P.~Marquard, J.~H.~Piclum, D.~Seidel, M.~Steinhauser,
  {\it Completely automated computation of the heavy-fermion corrections to the
three-loop matching coefficient of the vector current},
  Phys.\ Lett.\  B {\bf 678 } (2009)  269
  [arXiv:0904.0920 [hep-ph]].

\bibitem{Anzai:2009tm}
  C.~Anzai, Y.~Kiyo, Y.~Sumino,
  {\it Static QCD potential at three-loop order},
  Phys.\ Rev.\ Lett.\  {\bf 104 } (2010)  112003
  [arXiv:0911.4335 [hep-ph]].

\bibitem{Smirnov:2009fh}
  A.~V.~Smirnov, V.~A.~Smirnov, M.~Steinhauser,
  {\it Three-loop static potential},
  Phys.\ Rev.\ Lett.\  {\bf 104 } (2010)  112002
  [arXiv:0911.4742 [hep-ph]].

\bibitem{Hoang:2000ib}
  A.~H.~Hoang, A.~V.~Manohar, I.~W.~Stewart, T.~Teubner,
  {\it A Renormalization group improved calculation of top quark production near
threshold},
  Phys.\ Rev.\ Lett.\  {\bf 86 } (2001)  1951
  [hep-ph/0011254].

\bibitem{Penin:2004ay}
  A.~A.~Penin, A.~Pineda, V.~A.~Smirnov, M.~Steinhauser,
  {\it Spin dependence of heavy quarkonium production and annihilation rates:
Complete next-to-next-to-leading logarithmic result},
  Nucl.\ Phys.\  B {\bf 699 } (2004)  183
  [hep-ph/0406175].

\bibitem{Pineda:2006ri}
  A.~Pineda, A.~Signer,
  {\it Heavy Quark Pair Production near Threshold with Potential
Non-Relativistic QCD},
  Nucl.\ Phys.\  B {\bf 762 } (2007)  67
  [hep-ph/0607239].

\bibitem{Bigi:1991mi}
  I.~I.~Y.~Bigi, V.~S.~Fadin, V.~A.~Khoze,
  {\it Stop near threshold},
  Nucl.\ Phys.\  B {\bf 377 } (1992)  461.

\bibitem{Melnikov:1993np}
  K.~Melnikov, O.~I.~Yakovlev,
  {\it Top near threshold: All $\alpha_s$ corrections are trivial},
  Phys.\ Lett.\  B {\bf 324 } (1994)  217
  [hep-ph/9302311].

\bibitem{Penin:1998ik}
  A.~A.~Penin, A.~A.~Pivovarov,
  {\it Top quark threshold production in $\gamma\gamma$ collision in the
next-to-leading order},
  Nucl.\ Phys.\  B {\bf 550 } (1999)  375
  [hep-ph/9810496].

\bibitem{Hoang:2004tg}
  A.~H.~Hoang, C.~J.~Rei{\ss}er,
  {\it Electroweak absorptive parts in NRQCD matching conditions},
  Phys.\ Rev.\  D {\bf 71 } (2005)  074022
  [hep-ph/0412258].

\bibitem{Hoang:2010gu}
  A.~H.~Hoang, C.~J.~Rei{\ss}er, P.~Ruiz-Femen\'{i}a,
  {\it Phase Space Matching and Finite Lifetime Effects for Top-Pair Production
Close to Threshold},
  Phys.\ Rev.\  D {\bf 82 } (2010)  014005
  [arXiv:1002.3223 [hep-ph]].

\bibitem{Beneke:2010mp}
  M.~Beneke, B.~Jantzen, P.~Ruiz-Femen\'{i}a,
  {\it Electroweak non-resonant NLO corrections to $e^+ e^- \to W^+ W^- b
  \bar{b}$ in the $t \bar{t}$ resonance region},
  Nucl.\ Phys.\  B {\bf 840 } (2010)  186
  [arXiv:1004.2188 [hep-ph]].




\bibitem{Pineda:1997bj}
  A.~Pineda, J.~Soto,
  {\it Effective field theory for ultrasoft momenta in NRQCD and NRQED},
  Nucl.\ Phys.\ Proc.\ Suppl.\  {\bf 64 } (1998)  428
  [hep-ph/9707481].

\bibitem{Brambilla:1999xf}
  N.~Brambilla, A.~Pineda, J.~Soto, A.~Vairo,
  {\it Potential NRQCD: An Effective theory for heavy quarkonium},
  Nucl.\ Phys.\  B {\bf 566 } (2000)  275
  [hep-ph/9907240].

\bibitem{Beneke:1997zp}
  M.~Beneke, V.~A.~Smirnov,
  {\it Asymptotic expansion of Feynman integrals near threshold},
  Nucl.\ Phys.\  {\bf B522 } (1998)  321
  [hep-ph/9711391].

\bibitem{Smirnov:2002pj}
  V.~A.~Smirnov,
  {\it Applied asymptotic expansions in momenta and masses},
  Springer Tracts Mod.\ Phys.\  {\bf 177 } (2002)  1.


\end{thebibliography}
\end{document}